\documentclass[twocolumn,
superscriptaddress,
amsmath,amssymb,
prl, 
floatfix
]{revtex4-2}

\usepackage{graphicx}% Include figure files
\usepackage{dcolumn}% Align table columns on decimal point
\usepackage{bm}% bold math
\usepackage{hyperref}
\usepackage{multirow}
\usepackage{array}
\usepackage{xcolor}
\usepackage[normalem]{ulem}
\usepackage{comment}

\begin{document}
\title{Spontaneous elongation of 3D gastruloids from local cell polarity alignment}

\author{Richard D.J.G. Ho}
\thanks{Equal contribution}
\affiliation{Njord Centre, Department of Physics, University of Oslo, Oslo, 0371, Norway}

\author{Endre J. L. Mossige}
%\orcid{0000-0002-6929-2932}}
\thanks{Equal contribution}
\affiliation{RITMO Centre for Interdisciplinary Studies in Rhythm, Time and Motion, University of Oslo, Forskningsveien 3A, 0373 Oslo, Norway}
\affiliation{Njord Centre, Department of Physics, University of Oslo, Oslo, 0371, Norway}

\author{Sergei Ponomartcev}
\affiliation{Hybrid Technology Hub, Faculty of Medicine, University of Oslo, Sognsvannsveien 9, 0372 Oslo, Norway}

\author{Natalia Smirnova}
\affiliation{Hybrid Technology Hub, Faculty of Medicine, University of Oslo, Sognsvannsveien 9, 0372 Oslo, Norway}

\author{Xian Hu}
\affiliation{Centre for Cancer Cell Reprogramming, Faculty of Medicine, University of Oslo, Ullernchausséen 70, 0379 Oslo, Norway}

\author{Keqing Sunny Dai}
\affiliation{Core Facilities for Integrated Microscopy, Faculty of Health and Medical Sciences, 
University of Copenhagen, Nørre Allé 20,
DK-2200, Copenhagen N, Denmark}

\author{Stefan Krauss}
\affiliation{Hybrid Technology Hub, Faculty of Medicine, University of Oslo, Sognsvannsveien 9, 0372 Oslo, Norway}

\author{Dag Kristian Dysthe}
\affiliation{Njord Centre, Department of Physics, University of Oslo, Oslo, 0371, Norway}

\author{Luiza Angheluta}
\email{luiza.angheluta@fys.uio.no}
\affiliation{Njord Centre, Department of Physics, University of Oslo, Oslo, 0371, Norway}

\begin{abstract}
Gastruloids, as 3D stem cell aggregates model for early embryogenesis, provide a unique platform to study how collective cell dynamics drive tissue symmetry breaking and axial elongation. Using 3D light-sheet imaging, we show that a pulse of Chiron, a Wnt activator, induces coherent alignment of cell polarity during elongation. While nuclear elongation occurs with or without treatment, only Chiron-treated gastruloids exhibit quasi–long-range alignment of nuclear axes, linking cell polarity coherence to tissue-scale remodeling. A minimal physical model of polarized cells, incorporating alignment-dependent torques and polarity-mediated adhesion, reproduces symmetry breaking and elongation, demonstrating that local cell polarity alignment alone can drive tissue-scale convergence–extension flows.
\end{abstract}

\maketitle

\section{Introduction}
Tissue morphogenesis is fundamental to organ and embryo development. This process emerges from a complex interplay of coordinated cell proliferation, differentiation and migration orchestrated by mechanical interactions and biochemical signals such as morphogens~\cite{heisenberg2013forces}. Morphogens establish spatial gradients that provide positional information, guiding cells toward specific fates and ultimately shaping tissue architecture~\cite{rogers2011morphogen}.

Gastruloids are powerful in-vitro models that capture several aspects of early embryogenesis, including polarization, axial elongation~\cite{beccari2018multi,simunovic2017embryoids}, and gastrulation-like events~\cite{martyn2018self}. 
As such, gastruloids are three-dimensional multicellular aggregates of stem cells that are induced with chemicals to follow specific embryo-like developmental paths. A key experimental perturbation involves treatment with Chir99021 (Chi), a small molecule activator of the canonical Wnt/$\beta$-catenin signaling pathway, which is added to the cell aggregate 2 days post aggregation (dpa) to induce tissue elongation. 
This axial elongation, a hallmark of gastruloid development, is a process that integrates genetic regulation, biochemical signaling, and mechanical forces~\cite{beccari2018multi, moris2020vitro}. 
Recent experimental evidence highlights the importance of both active cell crawling and differential adhesion in driving convergence-extension flows, which act as a cell sorting mechanism at the tissue scale. These processes generate a spectrum of gastruloid shapes, from uniaxial elongation to multi-axial and irregular forms~\cite{de2024shapes}. Notably, robust, uniaxial elongation and stable gene expression profiles form only in gastruloids made from an optimal initial cell number; outside this range, size-dependent effects lead to abnormal morphologies and altered cell fate patterns~\cite{fiuza2024morphogenetic}. Furthermore, recent experimental studies show that, through cell sorting, the initial heterogeneity in Wnt signaling results in the posterior localization of Wnt-high cells and polarization of the gastruloid~\cite{mcnamara2024recording}. 

Conventional modeling approaches to gastruloid elongation have primarily focused on reaction-diffusion mechanisms of morphogens, such as Wnt, BMP, and Nodal, to explain symmetry breaking and axis formation. These models, inspired by Turing’s theory, posit that self-organized spatial gradients of diffusible signals can drive pattern formation even in the absence of pre-existing spatial cues~\cite{turing1990chemical, mcnamara2024recording}. Indeed, recent studies show that gastruloids, when uniformly stimulated with Wnt signaling, can spontaneously develop a polarized domain of Wnt activity leading to uniaxial elongation~\cite{mcnamara2024recording}. While reaction-diffusion models successfully explain the emergence of biochemical patterns, they do not account for the mechanism of cell sorting.

Differential adhesion is the predominant explanation for cell sorting during morphogenesis~\cite{Steinberg2007} and the
cellular Potts model (CPM) is a commonly used computational model for simulating tissue organization driven by differential adhesion~\cite{glazier1993simulation,hogeweg2000evolving}. Extensions of the CPM have successfully modeled convergent extension, in which cells intercalate and elongate aggregates along a principal axis, driven by anisotropic adhesion and active cell behavior~\cite{zajac2003simulating,keller2000mechanisms}. However, while CPM-based models capture many aspects of morphogenesis, they often rely on predefined rules for cell movement and may not fully account for the intrinsic polarity and emergent collective behaviors observed in systems like gastruloids. 

%--------------------figure 1 ------------------------
\begin{figure}
    \centering
    \includegraphics[width=1.0\linewidth]{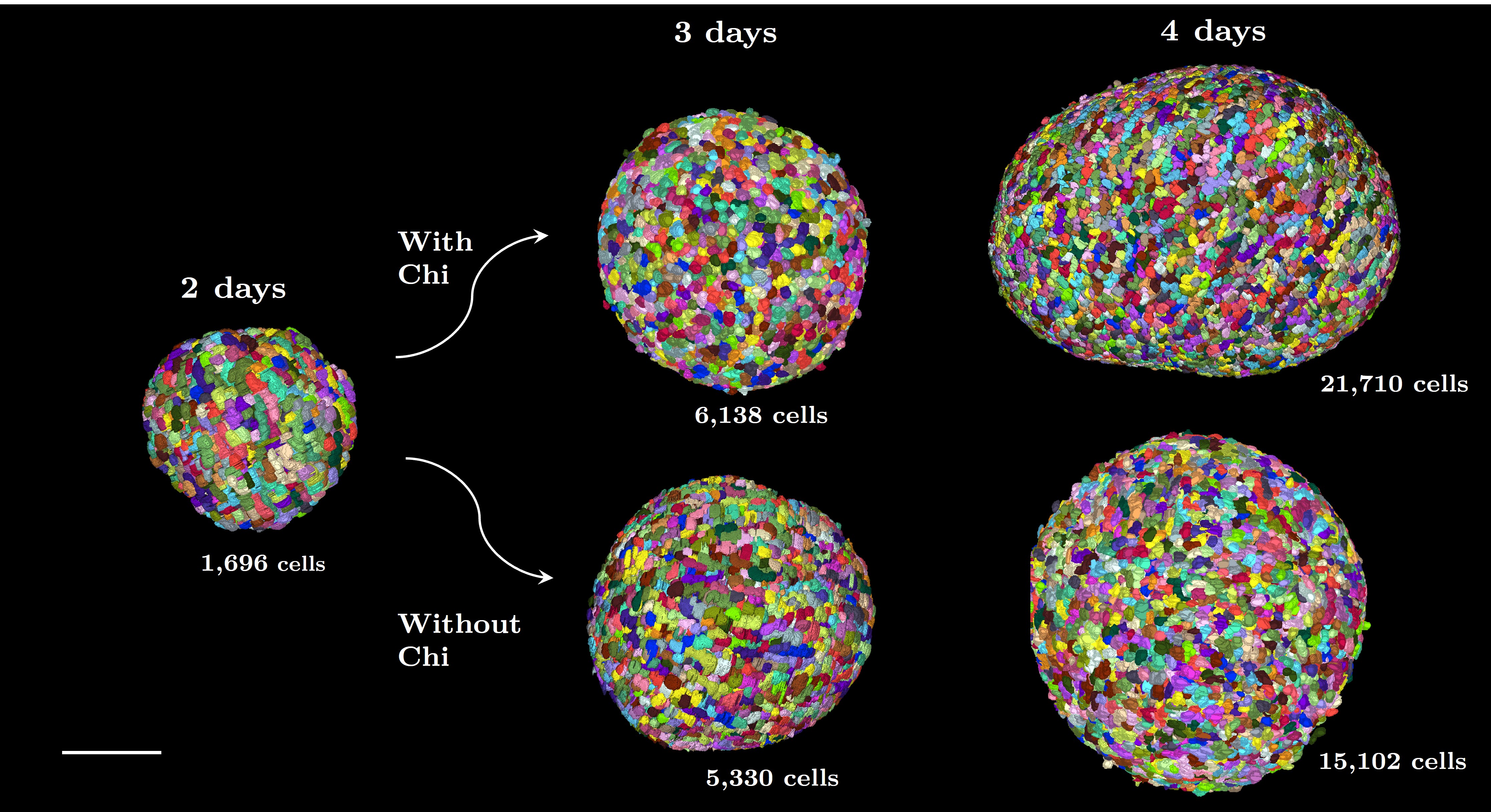}
    \caption{Shapes of representative color-coded gastruloid samples, obtained from 3D segmented light sheet image data. Top row shows gastruloids at 3 to 4 days after aggregation (dpa) with Chi, showing sign of elongation. Bottom row show control samples at the same time points without Chi. Scale bar 100 $\mu$m.  }
\label{fig:3DshapesExperiments}
\end{figure}

As an alternative approach, we employ a minimal model of interacting polarized cells, introduced in Refs.~\cite{nissen2018theoretical,nielsen2020model}, where cell polarities are coupled with cell migration, allowing for self-organization and morphogenetic events to emerge from local interaction forces. This approach has recently been shown to capture key morphological changes in epithelial sheets, including defect-mediated tissue curvature and tubulogenesis~\cite{ho2024role}, which are also prominent features in gastruloid development.

In this work, we integrate high-resolution experimental data from advanced 3D light sheet imaging of mouse gastruloids~\cite{van2014symmetry} with theoretical predictions from a polarized cell model to uncover the physical mechanisms driving early symmetry breaking during mammalian gastruloid development. Our imaging analysis reveals that, by day 3 post-aggregation, cell nuclei become elongated irrespective of Chi treatment. However, upon Chi addition, gastruloids exhibit long-range alignment of nuclear axes spanning the entire tissue, in stark contrast to control samples, which display only short-range orientational correlations without global order. A minimal model incorporating self-aligning torques on cell polarity and polarity-dependent adhesion forces quantitatively reproduces this transition, predicting collective cell migration and spontaneous tissue elongation consistent with experimental observations.

\section{Mouse gastruloids: from cell polarity to gastruloid polarization}

To explore whether and how cell polarity influences morphological changes in early gastruloids and what is the role of GSK3 $\beta$ inhibition/activation of canonical Wnt signaling in the process, we create mouse gastruloids based on the Beccari protocol~\cite{beccari2018multi}.  Gastruloids are allowed to develop for four days post-aggregation, either in the presence or absence of Chi. During this time window, gastruloids should undergo symmetry breaking and exhibit early features of axial development. For the analysis of the fixed specimen, we use light sheet fluorescence microscopy (Zeiss Lightsheet 7) combined with 3D image analysis at days 2, 3, and 4 after cell aggregation. We focus on comparing untreated gastruloids to those treated with Chi between day 2 and day 3 to study the effects of GSK3$\beta$ inhibition/activation of canonical Wnt signaling on nuclear deformation or elongation that we use as a proxy for tissue-level cellular polarization~\cite{Versaevel2012,Grosser2021}. Chi functions by inhibiting GSK3$\beta$, which i) stabilizes $\beta$-catenin, thereby triggering an activation of canonical Wnt signaling and impacting adherence junctions and ii) alters microtubule stability and actin dynamics. Hence, Chi treatment may likely influence cellular polarity, cell adhesion and by that the balance between isotropic and anisotropic distribution of cells in the developing gastruloids that naturally emerges through self-organization.

To extract cell polarities from 3D imaging, we ease the task of segmentation by fluorescent labeling the nuclei with SYTOX Deep Red.  When a cell is polarized due to external constraints or during migration, the cytoskeleton deforms the nucleus to an elongated shape similar to that of the cell itself~\cite{Versaevel2012,Grosser2021}. After fixation and staining, we render the samples transparent using the EZ Clear protocol~\cite{hsu2022ez} to enable deep tissue imaging with near-infrared light ($647$ nm), a method that reduces scattering and improves image quality compared to conventional UV dyes. Subsequently, image stacks from both sides of each gastruloid are de-convolved and fused (Huygens Professional) for optimal 3D reconstruction. Nuclear segmentation is performed in 3D using Cellpose (with the pretrained model "Cyto")~\cite{stringer2021cellpose}, and custom Python scripts extracted orientation, size, and volume from the binary masks ~\cite{Mossige_Wilson_Hu_Ho_Røberg-Larsen_Angheluta-Bauer_Krauss_Kjos_Jensenius_Dysthe_etal._2025}. To ensure that clearing and staining does not distort morphology, we also image uncleared, unstained gastruloids using a custom-built light sheet microscope~\cite{OsloSPIM,WhoHasAnOpenSPIM}.

\begin{figure}
    \centering
\includegraphics[width=1.0\linewidth]{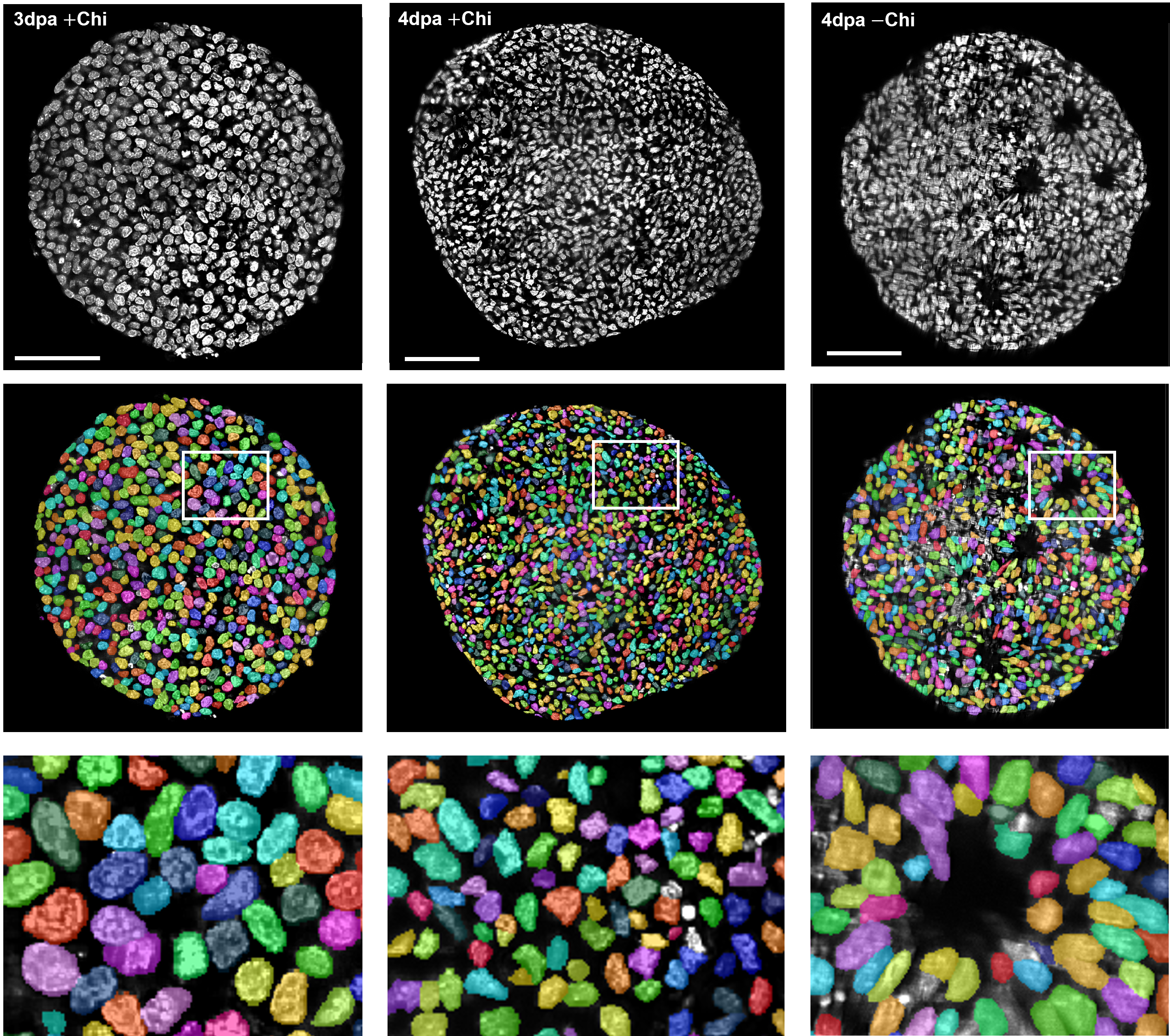}
    \caption{\textbf{Top row:} Representative optical cross sections of gastruloids at 3 dpa with ($+$) Chi (left), 4 dpa with ($+$) Chi (middle), and 4 dpa without ($-$) Chi (right). \textbf{Middle row:} 2D labeled masks after segmentation, overlaid the images above. \textbf{\textbf{Bottom row:}} Enlarged views of the segmented nuclei inside the marked regions show that nuclei for 3 dpa gastruloids exposed to Chi (left) appear more rounded and less elongated compared to samples without exposure to Chi. The rightmost figure shows elongated nuclei organized around a presumed emerging neural rosette. Scale bars 100 $\mu$m.}
\label{fig:2DSlicesChironExperimentsLabels}
\end{figure}

Figure~\ref{fig:3DshapesExperiments} shows 3D shapes of representative gastruloids at successive days with and without Chi, obtained from 3D segmented data. While gastruloids treated with Chi (top row) form spheroids at day 3 and show elongation at day $4$, control aggregates not treated with Chi maintain their round shape without clear signs of elongating. The top row of Figure ~\ref{fig:2DSlicesChironExperimentsLabels} shows typical single-plane optical sections of gastruloids at days $3$ and $4$, with ($+$) and without ($-$) Chi. The middle and bottom rows show color coded 2D masks after segmentation.

%---------------figure ------------------------
%---------------end figure ------------------------

To quantify nuclear orientation as a proxy for cell polarity, we apply principal component analysis (PCA) to segmented ellipsoidal nuclei. For each nucleus $i$, PCA yields three unit eigenvectors $\{\mathbf{e}_1^{(i)}, \mathbf{e}_2^{(i)},\mathbf{e}_3^{(i)}\}$, defining the principal axes of nuclear shape (up to a sign), along with corresponding eigenvalues that represent the extent of elongation along each axis. Nuclear elongation is quantified by the eigenvalue ratio, given  by the ratio of the largest to smallest eigenvalue. Outliers with large nuclear volumes or excessive elongation may result from segmentation artifacts which could bias the analysis of polarity alignment. Therefore, to ensure reliable orientation measurements, we consider only nuclei with volumes between $65$ and $4189 \, \mu m^3$  (equivalent to diameters of $5–20\,\mu m$) and an eigenvalue ratio below $2.5$. 

\begin{figure}
\centering  \includegraphics[width=0.95\linewidth]{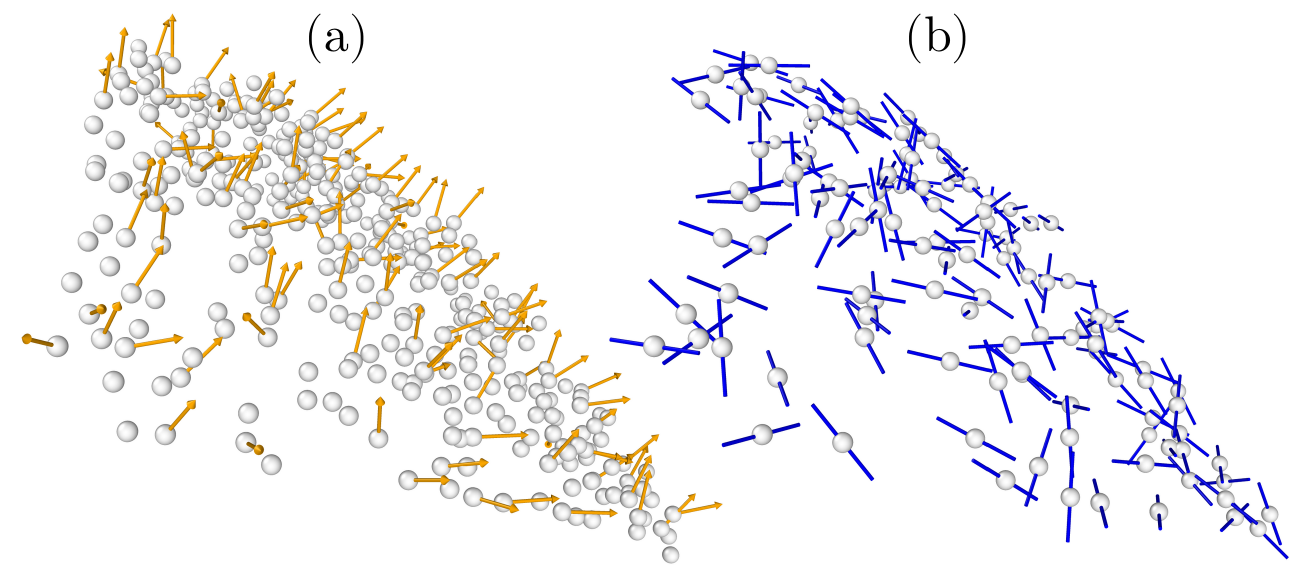}
  \caption{\textbf{Representation of principal nuclear orientations for cells on the surface of a representative 4 dpa +Chi gastruloid.} Spheres represent the centroid of nuclei identified as on the surface: (a) $\mathbf e_3$ eigenvectors along the shortest axis (orange) with the direction pointing away from the surface; (b) $\mathbf e_1$ nematic directors along the longest axis (blue), these are 40$\mu m$ in length to give a sense of scale. Only a third of nuclei on the surface have their eigenvector shown, whereas only a third of nuclei are shown for the nematic directors, both for clarity.}
  \label{fig:example_surface_polarity}
\end{figure}

Figure~\ref{fig:example_surface_polarity} illustrates the nuclear orientation vectors corresponding to the cell nuclei on the surface layers of a gastruloid at 4 dpa with Chi.  The shortest axes $\mathbf e_3$ typically align perpendicular to the surface, consistent with an apical-basal polarity. In contrast, the longest axes $\mathbf e_1$ tend to lie within the surface tissue, similar to planar cell polarities. These surface effects are also seen in gastruloids not treated with Chi.

To quantify the orientational alignment of nuclear polarities, we analyze the radial correlations of orientation vectors as function of their separation distance defined as
\begin{align}
C_n (r) 
= \frac{1}{N_{p}(r)}
    \sum\limits_{i} \sum\limits_{\substack{j \\ r - \delta r \leq |\mathbf{r}_{ij}| \le r + \delta r}} 
    \left| \mathbf{e}_n^{(i)} \cdot \mathbf{e}_n^{(j)} \right|,
\end{align}
where $\mathbf r_{ij} = \mathbf r_i-\mathbf r_j$ is the distance vector between nuclei $i$ and $j$, $N_p(r)$ is the number of pairs that contribute to each radius bin, and $n = 1,2,3$ is the eigenvector label. Notice the absolute value of the inner product because the eigenvectors are defined up to arbitrary signs. Unlike the conventional pair correlation for spin systems, the asymptotic limit of $C_n(r)$ as $r\rightarrow \infty$ is not a measure of global average orientation $|\langle\mathbf e_n\rangle|^2$, but instead it is given by $C_n^\infty = \lim\limits_{r\rightarrow\infty}C_n(r)= \langle|\cos(\theta)|\rangle = 1/2$ in 3D, where $\theta$ is the angle between two randomly orientated unit vectors. We subtract this baseline correlation from $C_n(r)$ and normalize it as $\Delta C_n(r) = \frac{C_n - C_n^\infty}{1-C_n^\infty}$, such that at zero distance it equals $1$. 
This analysis is appropriate for infinite systems which have zero correlation between distant points. For the numerical model, due to the strong finite size effects, we instead use $C_n^\infty = N^{-2}\sum\limits_{ij} | \mathbf{e}_n^{(i)} \cdot \mathbf{e}_n^{(j)}|$, which deviates slightly from the asymptotic value $1/2$ for infinite systems. 
%--------------- figure 3-----------------
%-----------------------------------

\begin{figure}
\centering
\includegraphics[width=0.95\linewidth]{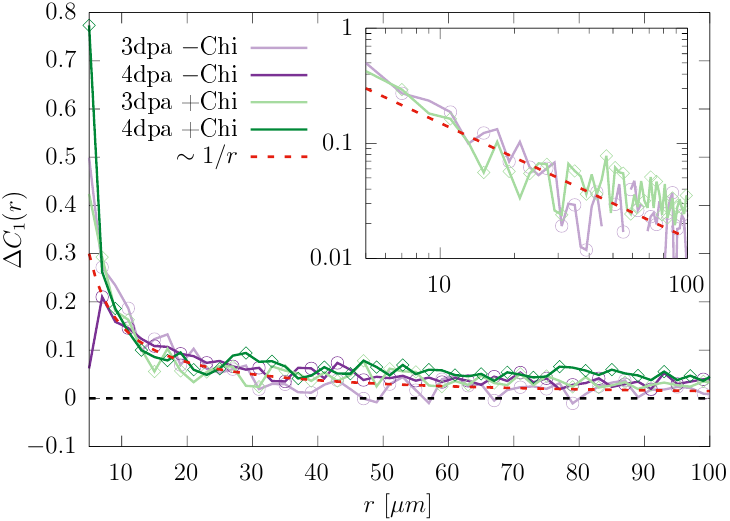}
  \caption{{\bf Experimentally measured pair correlations on the surface:} Pair correlation function of $\mathbf e_1$ orientations of longest axes at 3 dpa $\pm$ Chi and 4 dpa $\pm $ Chi on the gastruloid surface. Inset panel: log-log plot showing persistent orientational correlations on the gastruloid surface with or without Chi exposure. }
  \label{fig:surface_polarity}
\end{figure}

To quantify spatial patterns in the nuclear orientation on the gastruloid surfaces at 3 and 4 dpa \emph{without Chi}, and at 3 and 4 dpa \emph{with Chi}, we compute the pair correlation function $\Delta C_1(r)$, corresponding to the longest nuclear axes as a measure of planar cell polarity. This pair correlation has monotonic dependence on large in-surface distance as $\Delta C_1(r)\sim 1/r$ with an algebraic decay as shown in Figure~\ref{fig:surface_polarity}. This suggests that gastruloids develop persistent ordering in nuclear orientations on the surface with or without Chi exposure. 

\begin{figure}
\centering
\includegraphics[width=0.95\linewidth]{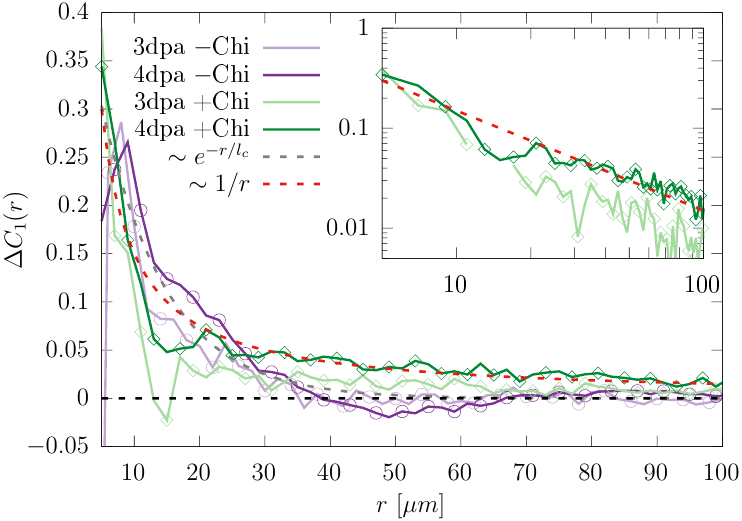}
  \caption{{\bf Experimentally measured pair correlations in the bulk:}  Pair correlation functions of nuclear orientations $\mathbf e_1$ in gastruloid bulk at 3 and 4 dpa with and without Chi. As inset is the log-log plot of $\Delta C_1$ for 3 and 4 dpa with Chi, showing the persistent bulk orientational correlations only in Chi-treated gastruloids. }
\label{fig:bulk_polarities}
\end{figure}

A similar analysis is performed for nuclei in the bulk at 3 and 4 dpa, as shown in Figure~\ref{fig:bulk_polarities}. In gastruloids \emph{without Chi}, the pair correlation $\Delta C_1(r)\sim e^{-r/r_c}$ tends to decay exponentially on short distances with a correlation length $l_c \approx 10\,\mu$m, corresponding to the mean nuclear spacing. At larger distances ($r>30 \mu$m), $\Delta C_1$ alternates sign, meaning that $C_1(r)$ occasionally dips below the average long-range order. These variations relative to the global order may reflect the distinct orientational domains, similar to those observed in ferromagnets or liquid crystals.  However, only the gastruloids \emph{with Chi} show persistent correlations at large distances as $\Delta C_1(r)\sim 1/r$ with an algebraic decay consistent with quasi-long-range order. These results indicate that Chi treatment promotes alignment of nuclear and cellular polarity as early as 3-4 days post aggregation.

%-------------- figure 4 -------------
%-------------- figure -------------

%-------------- figure 5 -------------

\section{Cell polarity model: Tissue elongation from alignment torque on polarity and cell migration}
To increase our understanding of the processes that lead to the observed morpho-types in the experimental gastruloids, we consider a minimal framework where each cell is represented as a point particle endowed with two polarity vectors: $\mathbf{p}$ representing planar cell polarity (PCP) and $\mathbf{q}$ representing apical-basal (AB) polarity~\cite{nissen2018theoretical}. The PCP $\mathbf p$ encodes anisotropic cell orientations resulting from various mechanisms including biochemical regulations such as non-canonical Wnt signaling pathways~\cite{gao2011wnt}. This biochemical feedback is effectively modeled as a polarity torque that promotes alignment of a cell’s polarity with those of its neighboring cells. 

Cells interact with their nearest neighbors through a pairwise potential $V_{ij} = V_{ij}^{(0)} + V_{ij}^{(1)}$, where $V_{ij}^{(0)}$ is an isotropic soft potential for dense fluids with short-range repulsion and long-range attraction (discussed more later), and $V_{ij}^{(1)}= e^{-r_{ij}} - (\lambda_1 S_1 + \lambda_2 S_2 + \lambda_3 S_3) e^{-r_{ij} / a_0}$ is anisotropic, modulated by the relative orientation of cell polarities and their separation vector~\cite{nissen2018theoretical}. The position of the $i$th particle is labeled as $\mathbf r_i$, while its pair separation vector from the $j$th particle is $\mathbf r_{ij} = \mathbf r_i-\mathbf r_j$, with  distance magnitude $r_{ij} = |\mathbf r_{ij}|$ and orientation vector $\hat{\mathbf r}_{ij} = \mathbf r_{ij}/r_{ij}$. The strength of anisotropic interactions is governed by parameters $\lambda_1$, $\lambda_2$, and $\lambda_3$, and the terms $S_1$, $S_2$, $S_3$ encode orientation alignment of polarities with respect to each other and to cell positions, 
\begin{eqnarray}
\begin{cases}
 S_1 = (\mathbf p_i \times \hat{\mathbf r}_{ij}) \cdot (\mathbf p_j \times \hat{\mathbf r}_{ij}) \\
S_2 = (\mathbf p_i \times \mathbf q_{i}) \cdot (\mathbf p_j \times \mathbf q_{j}) \\
S_3 = (\mathbf q_i \times \hat{\mathbf r}_{ij}) \cdot (\mathbf q_j \times \hat{\mathbf r}_{ij}),   
\end{cases}
\end{eqnarray}
such that $S_1=S_2=S_3=1$ for equilibrium configurations. These anisotropic interaction terms capture distinct states of cell polarity alignment versus spatial arrangements. The terms $S_1$ and $S_3$ promote alignment of the PCP $\mathbf{p}_i$ and AB polarity $\mathbf{q}_i$, respectively, that is orthogonal to the pair separation vector $\hat{\mathbf{r}}_{ij}$. In contrast, the $S_2$ term captures internal alignment between the PCP and AB axes regardless of their spatial positioning, i.e., it promotes alignment of the cross products $\mathbf{p}_i \times \mathbf{q}_i$ across neighboring cells. This cross product defines a normal vector to the $(\mathbf{p}_i, \mathbf{q}_i)$-plane and locally encodes the polarity handedness. 

We consider an overdamped dynamics of cell motion driven by gradient forces and translational noise  
\begin{eqnarray}\label{eq:migration}
 \dot{\mathbf{r}}_i = - \sum\limits_{j\in\partial i}\nabla_{\mathbf r_i} V_{ij}+\sqrt{2D_T}\boldsymbol{\xi}_i.   
\end{eqnarray}
Their polarities can reorient due to polarity-aligning torques induced by the neighboring cells or rotational noise, 
\begin{eqnarray}
\begin{cases}
  \dot{\mathbf{p}}_i = \left[\left(\sum\limits_{j\in\partial i}\nabla_{\mathbf p_i} V_{ij}\right)\times \mathbf p_i+\sqrt{2 D_R}\boldsymbol{\eta}_i\right]\times \mathbf p_i  \\
  \\
  \dot{\mathbf{q}}_i = \left[\left(\sum\limits_{j\in\partial i}\nabla_{\mathbf q_i} V_{ij}\right)\times \mathbf q_i+\sqrt{2 D_R}\boldsymbol{\zeta}_i\right]\times \mathbf q_i.
  \end{cases}
\end{eqnarray}
We also include small uncorrelated noise that is Gaussian distributed with zero mean and standard deviation given by translational diffusivity $D_T$ for $\boldsymbol{\xi}_i$ and rotational diffusivity $D_R$ for $\boldsymbol{\eta}_i,\boldsymbol{\zeta}_i$ for each particle $i$. 

We model gastruloid elongation driven solely by cell polarity, holding cell number constant and excluding growth from cell division. Elongation still occurs without proliferation, indicating that cell division is not essential for the core morphogenetic process, though it may modulate its extent or timing. Incorporating division and multiple cell types is feasible but would greatly increase computational cost, as gastruloid volume triples over the two-day simulation; these model extensions are left for future work.

%-----------figure ----------------------
\begin{figure}
\centering
\includegraphics[width=0.9\linewidth]{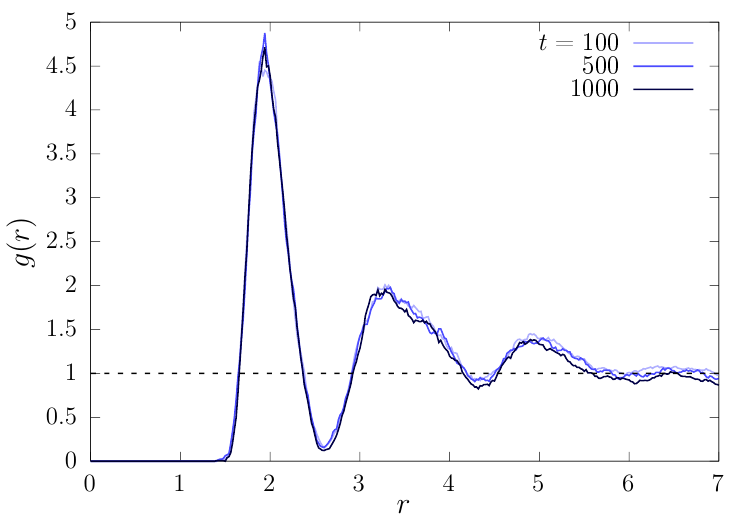}
  \caption{{\bf Model simulations of pair correlation in spatial arrangements.} Radial density function $g(r)$ for cells during elongation.}
  \label{fig:sims}
\end{figure}
%-----------figure ----------------------

To prevent this polarity-driven elongation from producing unrealistic cell arrangements, we introduce a crowding rule that mimics an effective surface tension: if a cell has fewer than a threshold number of neighbors, it only experiences the isotropic interaction potential
\begin{equation}
    V_{ij}^{(0)} = \lambda_0 \left( e^{-r_{ij}} - e^{-r_{ij}/\beta} \right).
\end{equation}
Without this rule, when $\lambda_1 = \lambda_3$ the system tends to form strings of cells regardless of $\lambda_2$, which is undesirable. Cells migrate following Eq.~\ref{eq:migration} with nearest neighbors determined using a local geometric rule: two cells are neighbors if they remain each other’s closest partners relative to all others. When another cell comes close enough to disrupt this pairing, a neighbor exchange occurs, resulting in relative cell migration and localized tissue fluidization \cite{ho2024role}.  

We first probe the spatial organization of cells within the aggregate by measuring the radial density profile $g(r)$, defined as the average number of neighbors at distance $r$ from a reference cell, normalized by the expectation for a uniform distribution. The resulting profile of $g(r)$ exhibits the characteristic structure of a dense fluid (Fig.~\ref{fig:sims}) and is robust across model parameters.

\begin{figure}
\centering
\includegraphics[width=1\linewidth]{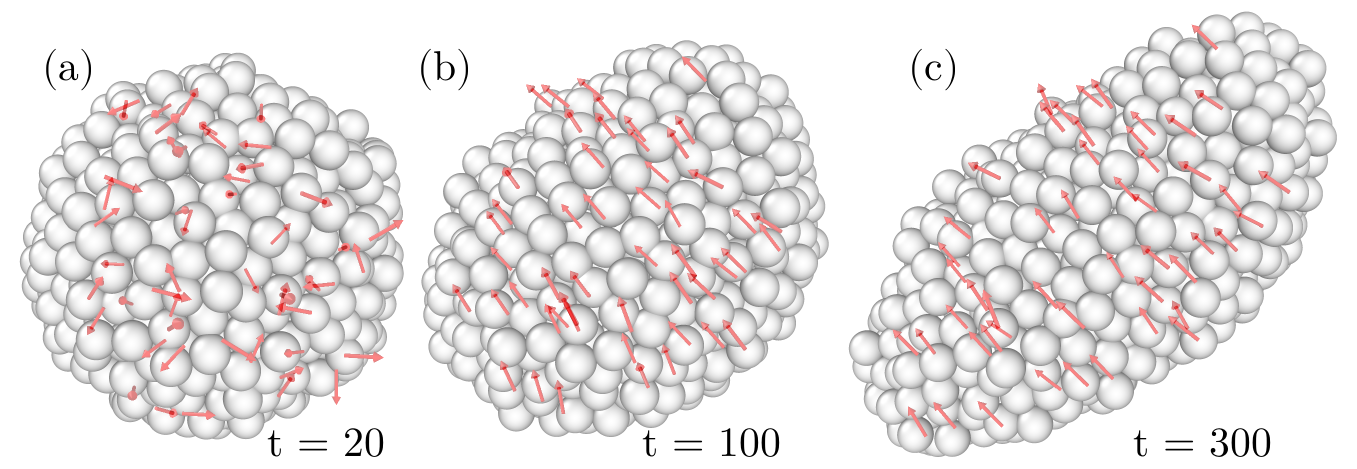}
\includegraphics[width=0.5\linewidth]{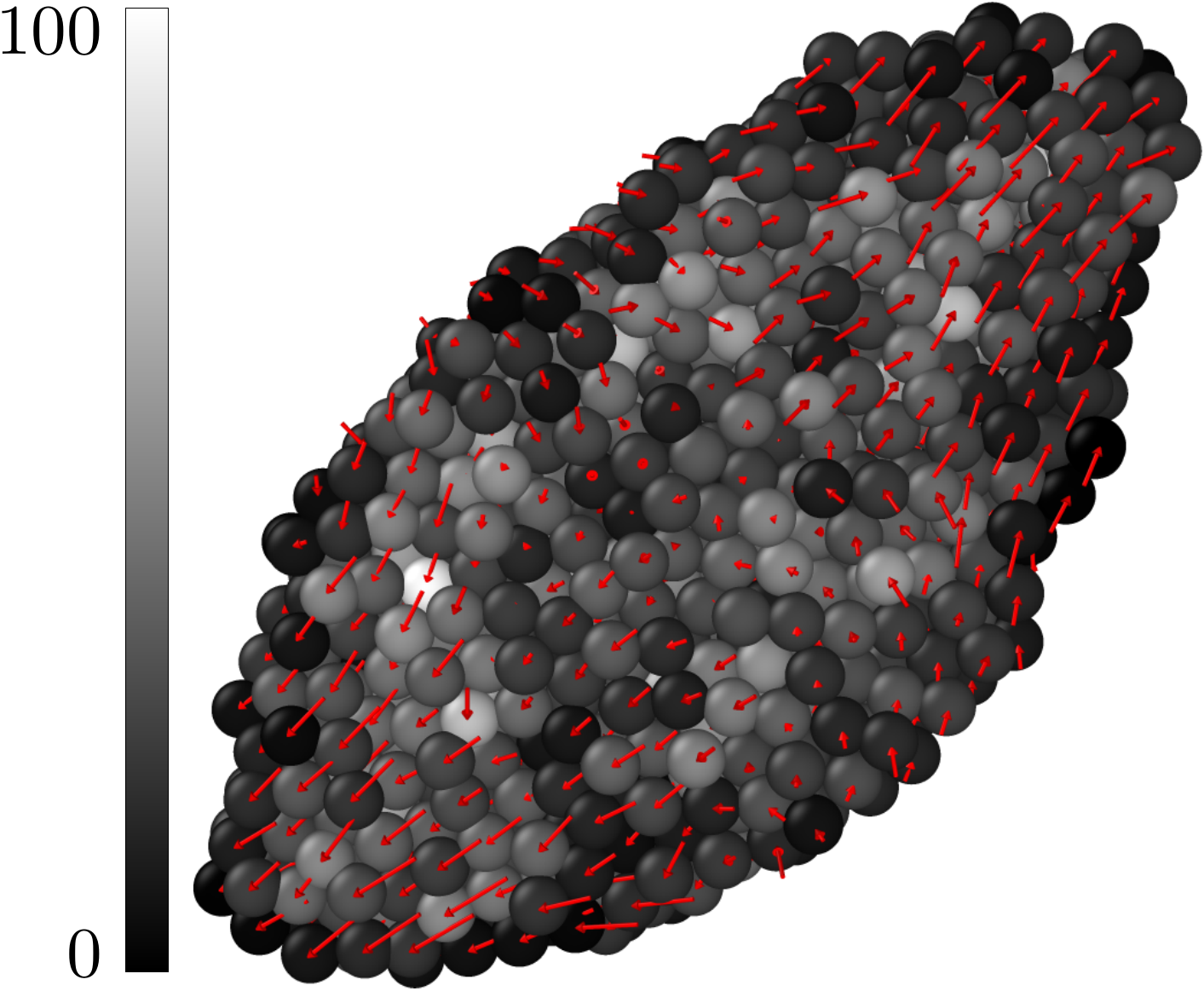}
   \caption{{\bf Model Simulations of PCP alignment and convergence-extension flow.} (Top) Time snapshots of cell aggregates during elongation. Red arrows represent the PCP polarity orientations. At early times, (a), the PCP orientations are more disordered, while during elongation, (b) and (c), the PCP polarities consistently align along a direction perpendicular to the elongation axis. (Bottom) Red arrows represent cell movement leading to convergence-extension flow patterns.  Grayscale gradient represents the number of neighbor exchanges a cell has undergone, with lighter color indicating more neighbor exchanges.}
  \label{fig:sims_gastruloids}
\end{figure}

Starting from a spherical aggregate with randomly oriented polarities, cells self-organize such that PCP polarities gradually align their orientation while the aggregate elongates perpendicular to this alignment axis (Fig.~\ref{fig:sims_gastruloids}). In our model, polarity alignment arises from two distinct mechanisms: (i) spatially-dependent torques (associated with $\lambda_{1,3}$), which bias the polarity axes $\mathbf{p}$ and $\mathbf{q}$ to lie in planes orthogonal to the intercellular vector $\hat{\mathbf r}$; and (ii) handedness-dependent torques (associated with $\lambda_{2}$), which favor $\mathbf{p}$ and $\mathbf{q}$ perpendicular to each other. By tuning $\lambda_{1,2,3}$ relative to each other and rotational noise, we explore the interplay between these aligning mechanisms and identify distinct regimes of tissue elongation.

%----------figure ---------------

\begin{figure}
\centering
\includegraphics[width=0.95\linewidth]{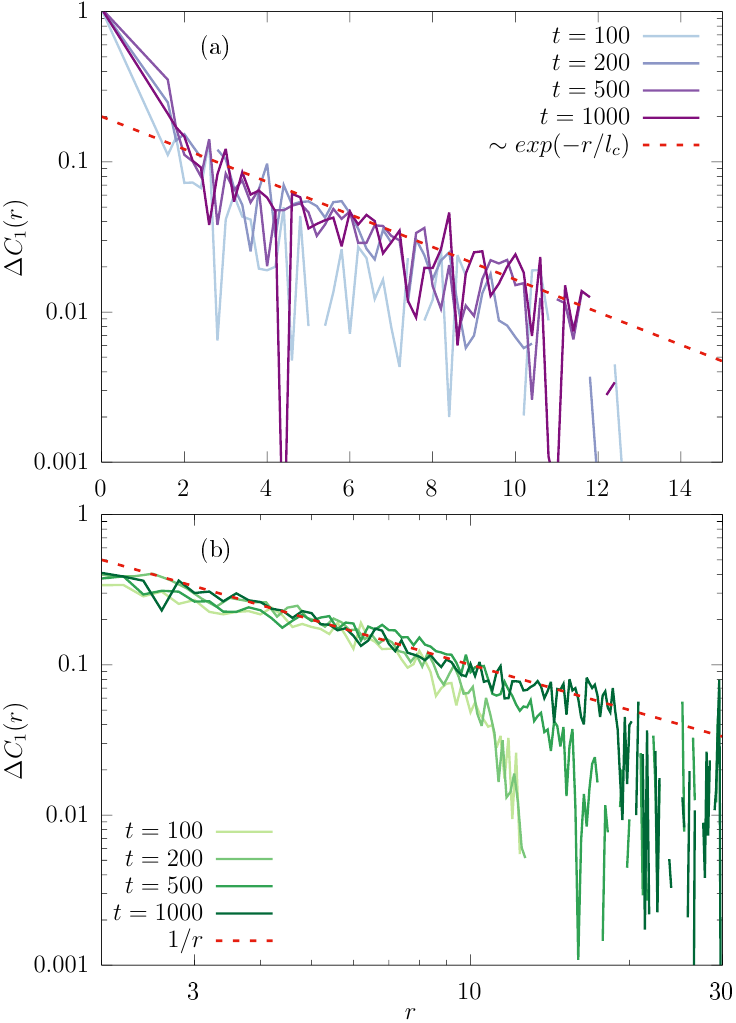}
  \caption{{\bf Model simulations of PCP correlation.} Relative pair correlation of PCP orientations for different times and for two regimes: (a) $\lambda_2 > \lambda_{1,3}$; (b) $\lambda_2 < \lambda_{1,3}$. 
  }
  \label{fig:sim_correlations}
\end{figure}

The pair-correlation of PCP polarity orientations $\Delta C_1(r)$ reflects these regimes (Fig.\ref{fig:sim_correlations}). When handedness-dependent torques dominate, correlations decay exponentially, indicating short-ranged alignment (Fig.\ref{fig:sim_correlations}a). Conversely, when spatially-dependent torques are the dominant interaction, they enforce alignment of neighboring cells’ polarity planes perpendicular to their distance vector $\hat{\mathbf r}$, thereby promoting coherent configurations characterized by quasi-long-range orientational order.(Fig.~\ref{fig:sim_correlations}b).

Despite these differences, both regimes support convergence–extension flows: cells migrate inward along the lateral sides and outward at the poles, thereby extending the anterior–posterior axis (Fig.~\ref{fig:sims_gastruloids}). Elongation is thus a bulk-driven process, arising from interior cells through local neighbor interactions and neighbor exchanges, while surface polarities remain largely disordered.

Tissue remodeling sustaining these flows is quantified by tracking cell neighbor exchanges over time. The resulting density map (Fig.~\ref{fig:sims_gastruloids}) shows that rearrangements concentrate in the interior, where fluidization is strongest, rather than at the surface. This implies that elongation can emerge from bulk dynamics alone, without the need for specialized leader cells at the boundary, though their presence in real tissues cannot be excluded.

%---------figure-----------

%----------------------
\section{Discussion and conclusions}
We show that local cell polarity alignment is sufficient to drive spontaneous elongation in 3D cell aggregates. Using light sheet microscopy and quantitative nuclear shape analysis, we demonstrate that Chi-treated mouse gastruloids exhibit persistent nuclear orientational alignment compared to untreated controls. Nuclear anisotropy, used as a proxy for cell polarity, reveal that nuclei in the outer layer orient their shortest axis perpendicular to the surface (consistent with apical–basal polarity), while their longest axis lies within the surface, resembling planar cell polarity (PCP).

We perform a spatial correlation analysis of the longest nuclear axes to explore the alignment of nuclei orientations. In Chi-treated gastruloids, correlations decay algebraically as $1/r$ in both surface and interior regions, indicating quasi–long-range order. By contrast, untreated controls exhibit  short-range correlations with exponential decay. These results suggest that Chi treatment promotes coherent polarity alignment across the tissue, coinciding with robust axial elongation.

These experimental results are further interpreted within the framework of a minimal theoretical model of polarized cells interacting through local adhesion and polarity alignment torques. The model includes two polarity axes, namely apical–basal ($\mathbf p$) and PCP ($\mathbf q$), and two polarity alignment mechanisms: spatially dependent torques ($\lambda_{1,3}$), which orient polarities perpendicular to the intercellular separation vector, and handedness-dependent torques ($\lambda_2$), which enforce perpendicularity between $\mathbf p$ and $\mathbf q$. While both mechanisms induce convergence-extension flows that result in tissue elongation, the emergence of quasi–long-range order in PCP orientation arises from the interplay between polarity alignment and spatial arrangements of cells via the spatially-dependent torque. 
Experimentally, the Chi treatment appears to enhance this alignment pathway, yielding coherent polarity and robust axial elongation. 

Despite its minimalism, our model provides a quantitative bridge between local cell-cell interactions and tissue-scale organization, demonstrating that polarity alignment through cell-cell interactions is sufficient to drive tissue-scale morphogenetic events.

More broadly, we propose an alternative approach to  morphogenesis based on intrinsic cell polarity alignment as driver for morphogenetic transitions in initially isotropic cell aggregates such as organoids, gastruloids, embryos, and other self-organizing tissues. Future model extensions to include self-propulsion or proliferation would further improve the range of model applicability to more diverse tissue developmental regimes.

\section*{Data availability}
Computer codes and image data is available in the OSF-repository found in Ref. \cite{Mossige_Wilson_Hu_Ho_Røberg-Larsen_Angheluta-Bauer_Krauss_Kjos_Jensenius_Dysthe_etal._2025}.

\section*{Acknowledgments} We thank Øyvind Ødegård Fougner and Ingrid Kjos for help with image processing, and Jana Harizanova at the Core Facility for Integrated Microscopy, University of Copenhagen for sample preparation, light sheet imaging,  image post processing, and for training EJLM.  We also thank the NorMIC Imaging Platform at the Department of Biosciences, University of Oslo, for providing assistance and access to the confocal microscope for quality control of gastruloids, as well as for the Cellpose/Python GPU image processing workstation.

This project was funded by the UiO:LifeScience project "Integrated Technologies for Tracking Organoid Morphogenesis", and three Norwegian centres of excellence: PoreLab, Hybrid Technology Hub, and RITMO Centre for Interdisciplinary Studies in Rhythm, Time and Motion. SK acknowledges support from the European Innovation Council (Pathfinder Grant, Supervised Morphogenesis in Gastruloids, Grant No. 101071203). EJLM also received funding from UiO Growth House, Kristine Bonnevie's reisestipend, and from the Bridging Nordic Microscopy Infrastructure Short-Term Scientific Missions program.

\bibliography{refs}
\end{document}